\documentclass[
 aip,
 jap,%
 amsmath,amssymb,
 reprint,%
]{revtex4-1}
\usepackage[final]{changes}
\usepackage{graphicx}
\usepackage{chngcntr}
\counterwithout{figure}{section}
\usepackage{dcolumn}
\usepackage{bm}
\usepackage{circuitikz}
\usepackage{tikz}
\usepackage{comment}
\newcommand{\A}{\mathcal{A}}
\newcommand{\D}{\mathcal{D}}
\newcommand{\G}{\mathcal{G}}

\newcommand{\M}{\mathcal{M}}
\newcommand{\B}{\mathcal{B}}

\newcommand{\J}{\mathcal{J}}
\newcommand{\F}{\mathcal{F}}
\newcommand{\K}{\mathcal{K}}

\begin{document}


\title{An Electromagnetic Induced Transparency-like Scheme for Wireless Power Transfer Using Dielectric Resonators}

\author{Sameh. Y. Elnaggar}
\email{s.elnaggar@unsw.edu.au.}
\affiliation{School of Engineering and Information Technology, University of New South Wales, Canberra, Australia.}

\date{\today}

\begin{abstract}
Similar to the hybridization of three atoms, three coupled resonators interact to form  bonding, anti-bonding and non-bonding modes. The non-bonding mode enables an electromagnetic induced transparency like transfer of energy. Here the non-bonding mode, resulting from the strong electric coupling of two dielectric resonators and an enclosure, is exploited to show that it is feasible to transfer power over a distance comparable to the operating wavelength. In this scheme, the enclosure acts as a mediator. The strong coupling permits the excitation of the non-bonding mode with high purity. This approach is different from resonant inductive coupling which works in the sub-wavelength regime. Optimal loads and the corresponding maximum efficiency are determined using two independent methods: Coupled Mode Theory and Circuit modelling. It is shown that, unlike resonant inductive coupling, the figure of merit depends on the enclosure quality and not on the load, which emphasizes the role of the enclosure as a mediator. Briefly after the input excitation is turned on, the energy in the receiver builds up via all coupled and spurious modes. As time elapses, all modes except the non-bonding cease to sustain. Due to the strong coupling between the dielectrics and the enclosure, such systems have unique properties such as high and uniform efficiency over large distances; and minimal fringing fields. These properties suggest that electromagnetic induced transparency like schemes which rely on the use of dielectric resonators can be used to power autonomous systems inside an enclosure or find applications when exposure to the fields needs to be minimal. Finite Element computations are used to verify the theoretical predictions by determining the transfer efficiency, fields profile and coupling coefficients for two different systems. It is shown that the three resonators must be present for efficient power transfer; if one or more are removed, the transfer efficiency reduces significantly.
\end{abstract}


\maketitle




\section{Introduction}
Since its introduction a decade ago, \cite{MIT_Science} Wireless Power Transfer (WPT) via resonant inductive coupling has attracted immense interest; this is mainly due to its significant efficiency over mid range distances. The efficiency is a function of both the coupling coefficient and the quality factors of the source and receiver resonators. In the past decade different high \emph{Q} resonant coils were designed and fabricated. In general, coupling falls off rapidly as the source and receiver are separated by a distance comparable to the coils diameter, hence limiting the transfer distance.\cite{SampleAPL} The insertion of a relay resonator between the source and receiver was shown to extend the transfer distance.\cite{Hamam2009, Zhang2012} The relay, source and receiver couple to form three coupled modes; one mode of interest, known in Ref. \cite{Hamam2009} as the \emph{dark} mode has the intriguing property of null fields in the relay. It is responsible for enhancing the transfer efficiency over an extended range by creating an Electromagnetic Induced Transparency like process (EIT like). However for most practical cases, the excitation of the dark mode without exciting the other two modes can be quite challenging.

On the other hand although dielectric resonators (DRs) appeared as a promising realization of WPT resonators in a very early work,\cite{MIT_Theory} it was later abandoned in favour of capacitively loaded coils. This might be partly due to the easiness of fabricating capacitively loaded loops where the electric field is confined within the capacitive region and the magnetic field extends well outside the loop to link with other resonators; and hence realizes inductive coupling. Inductive coupling does not cause the same safety hazards that electric coupling does. Recently, the use of DRs modes was revived.\cite{quadrupole2016} In this particular work, the high \emph{Q} quadrupole modes of two spherical DRs were exploited to realize an efficient WPT system.

Due to their low loss and compactness, DRs are used to improve the efficiency of Electron Spin Resonance Probes.\cite{Paper6} It was shown that a DR can strongly interact with an enclosing cavity; such interaction depends on the dielectric constant and the overlap areas.\cite{Paper4, Paper5} Unlike inductive coupling, this type of interaction is purely electric. It was also shown that two DRs placed in a cavity exhibit a behaviour similar to how the source, receiver and relay interact.\cite{Mattar11}

In the current article, we exploit the strong DR/Enclosure interaction to create an EIT like system capable of efficiently transferring power. Power can be either transferred within an enclosed cavity over a distance of the order of $\lambda$, or over a wireless gap where, at steady state, both the electric and magnetic energies are substantially small. Unlike convential WPT systems, the coupling is mainly electric. The negligible fringing fields suggest that the scheme can be used in high power or biomedical applications, where the exposure to fields needs to  be minimized.

In section II, we present the theory of three coupled centres using two independent approaches: Coupled Mode Theory (CMT) and circuit analysis. Expressions for the optimum load and maximum efficiency are derived. The conditions at which EIT-like transmission is possible are obtained and discussed. Finally, the transient response of the fields is briefly examined. Section III presents the application of the theory to two systems. Extensive full wave analysis is used to demonstrate the EIT-like main features, such as strong coupling, extended transfer distance, high efficiency and negligible fringing fields.
\section{Theory}
\begin{figure}[!h]
  \centering
   \includegraphics[width=3.5in]{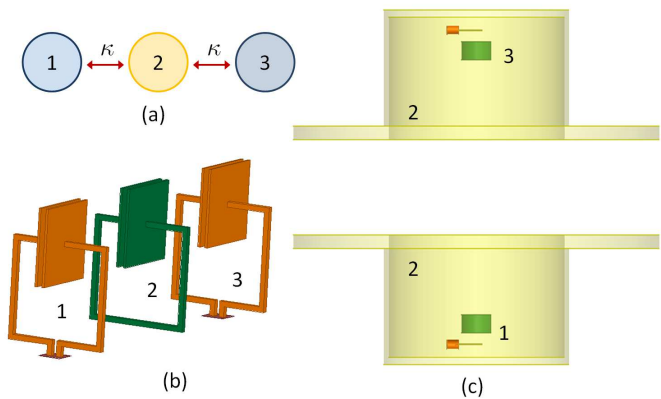}
   \caption{(a) Generic schematic of three coupled resonators. $\kappa$ is the coupling coefficient between resonator 1 (resonator 3) and resonator 2; there is no direct coupling between 1 and 3. (b) A system of three capacitively loaded loops.\cite{Stanford2011} (c) A 2DR/split cavity resonator system (2DR/SCR), the coupling between '1' and '3' is mediated via the SCR '2'.}
\label{fig:generic}
\end{figure}

Figures \ref{fig:generic}(b) and (c) depict two typical configurations that can be described using the coupling between the three generic centres shown in Fig. \ref{fig:generic}(a). If the separation between centres '1' and '3' is large, the direct interaction between them can be ignored. Nevertheless, centre '2', the mediator, still permits indirect interactions between them. The coupling between centre '1' ('3') and '2' is quantified by the coupling coefficient $\kappa$.\cite{Paper3} For simplicity, the coupling coefficient between the pairs ('1','2') and ('2','3') are taken to be equal. Unlike the capacitively loaded coils (Fig.\ref{fig:generic}b), the coupling between the DRs and enclosure modes in Fig.\ref{fig:generic}c is electric. Moreover due to the high dielectric constant and overlap areas, $\kappa$ can be substantially high.\cite{Paper4,Paper5}

The behaviour of the coupled system can be explained using CMT and/or circuit modelling. In the following subsections, the two approaches are used.
\subsection{Coupled Mode Theory}
In general, the interaction between three centres results in three coupled modes.\cite{Mattar11} Here we are mainly interested in the dark mode, hereafter called the non-bonding mode, since it resembles the non-bonding orbital in Molecular Orbital Theory. For the non-bonding mode, the enclosure/DR1  interaction is $180^\circ$ out of phase with the enclosure/DR3; resulting in a null component of the cavity mode. This property can be understood after one writes the eigenvalue problem of the coupled system.\cite{Paper2}
\begin{align}
\bm{\K}\bm{a} & =\omega^2 \bm{a} \label{eq:EigenvalueProb}\\
\left(\begin{array}{ccc}
\omega_0^2 &-\omega_2^2\kappa & 0\\
-\omega_0^2\kappa & \omega_2^2 & -\omega_0^2\kappa\\
0 &-\omega_2^2\kappa &\omega_0^2
\end{array} \right)
\left(\begin{array}{c}
a_1\\a_2\\a_3
\end{array} \right) & =\omega^2\left(\begin{array}{c}
a_1\\a_2\\a_3
\end{array} \right), \label{eq:EigenvalueProb2}
\end{align}
where $\omega_0$ is the resonant frequency of DR1 and DR3, $\omega_2$ is the frequency of the enclosure mode that strongly couples with the DR modes, $\omega$ and $\bm{a}$ are, respectively, the yet to be determined angular frequency and eigenvector of the coupled system.\cite{Paper2} In (\ref{eq:EigenvalueProb2}), the on diagonal terms responsible for   coupling induced frequency shifts are ignored.\cite{Popovic06} However, their effect is automatically included in the circuit model discussed in subsection B. It is readily seen that $\omega=\omega_0$, satisfies the secular equation $\det{(\bm{\K}-\omega^2\bm{I}})=\bm{0}$ and hence, is one of the eigenvalues of (\ref{eq:EigenvalueProb}). Moreover, the corresponding eigenvector is found to be
\begin{equation}
\label{eq:Eigenvector}
a=(-1,0,1)^\dagger
\end{equation}
As (\ref{eq:Eigenvector}) clearly shows, the non-bonding mode correlates the fields in DR1 to those in DR3 without exciting the enclosure fields. Unfortunately, it is extremely difficult to excite the non-bonding mode without exciting the other coupled and spurious modes. By careful design, spurious modes can be pushed away in frequency and thus minimizing their excitation. Similarly, a large value of $\kappa$ will push the frequency of the other two coupled modes away from the non-bonding mode and hence reduce their contribution to the excited fields. Toward this end, it will be assumed that excitation is applied to DR1 only. The steady state response $\tilde{\bm{a}}(\omega_\textnormal{in})$ to a sinusoidal input of frequency $\omega_\textnormal{in}$ is given by
\begin{align}
\label{eq:forcedECMT}
\tilde{\bm{a}}(\omega_\textnormal{in})=i\omega_\textnormal{in}\bm{\Phi}(\omega_\textnormal{in})\mathbf{J},
\end{align}
where $\bm{\Phi}(\omega_\textnormal{in})=\left(-\omega_\textnormal{in}^2\mathbf{I}+\bm{\K}\right)^{-1}$, $\bm{\K}$ here is the coupling matrix, given by (\ref{eq:EigenvalueProb}) but with the substitution of the complex frequencies $\gamma_k\equiv\omega_k+i|\sigma_k|$ in place of $\omega_k$ to take losses into account; $|\sigma_k|$ is the decay rate of the mode and it is equal to $\omega_k/2Q_k$, where $Q_k$ is the quality factor. $\mathbf{J}=\left(J_1~~0~~0\right)^T$ is the excitation vector \cite{URSI2016} (See also Appendix A). $\bm{\Phi}(\omega_\textnormal{in})$ is explicitly written as

\begin{equation*}
\bm{\Phi}(\omega_\textnormal{in})=\left(
\begin{array}{ccc}
-\omega_\textnormal{in}^2+\gamma_0^2 & -\gamma_2^2 \kappa & 0\\
-\gamma_0\kappa & -\omega_\textnormal{in}^2+\gamma_2^2 & -\gamma_3^2\kappa\\
0 & -\gamma_2^2\kappa & -\omega_\textnormal{in}^2+\gamma_3^2
\end{array}
\right)^{-1}
\end{equation*}

The ratios of the fields magnitudes $\tilde{a}_2(\omega_\textnormal{in})/\tilde{a}_1(\omega_\textnormal{in})$ and $\tilde{a}_3(\omega_\textnormal{in})/\tilde{a}_1(\omega_\textnormal{in})$ are determined from (\ref{eq:forcedECMT}) as
\begin{align*}
\frac{\tilde{a}_2(\omega_\textnormal{in})}{\tilde{a}_1(\omega_\textnormal{in})} & =\frac{\Phi_{21}}{\Phi_{11}},\\
\frac{\tilde{a}_3(\omega_\textnormal{in})}{\tilde{a}_1(\omega_\textnormal{in})} & =\frac{\Phi_{31}}{\Phi_{11}}.
\end{align*} 
The relevant $\bm{\Phi}$ elements can be found to be
\begin{align*}
\Phi_{11}\left(\omega_\textnormal{in}\right) & =\frac{1}{\det\bm{\Phi}}\left(-\omega_\textnormal{in}^2+\gamma_2^2\right)\left(-\omega_\textnormal{in}^2+\gamma_3^2\right)-\gamma_2^2\gamma_3^2\kappa^2,\\
\Phi_{21}(\omega_\textnormal{in}) & = \frac{1}{\det\bm{\Phi}}\gamma_2^2\left(-\omega_\textnormal{in}^2+\gamma_3^2\right)\kappa,\\
\Phi_{31}(\omega_\textnormal{in}) & = \frac{1}{\det\bm{\Phi}}\gamma_3^2\gamma_2^2\kappa^2.
\end{align*}

The ratio of the field's magnitudes in the enclosure mode to the DR1 (source) mode is then given by
\begin{align}
\label{eq:a2_a1}
\frac{\tilde{a}_2(\omega_\textnormal{in})}{\tilde{a}_1(\omega_\textnormal{in})} & =\frac{\gamma_2^2\left(-\omega_\textnormal{in}^2+\gamma_3^2\right)\kappa}{\left(-\omega_\textnormal{in}^2+\gamma_2^2\right)\left(-\omega_\textnormal{in}^2+\gamma_3^2\right)-\gamma_2^2\gamma_3^2\kappa^2}\\
& = \frac{\kappa\left[1-\left(\omega_\textnormal{in}/\gamma_3\right)^2\right]}{\left[1-\left(\omega_\textnormal{in}/\gamma_3\right)^2\right]\left[1-\left(\omega_\textnormal{in}/\gamma_2\right)^2\right]-\kappa^2}.
\end{align}

Hereafter, the frequency of the enclosure mode $\omega_2$ is taken to be equal to $\omega_0$ (i.e, the three resonators have the same resonant frequency $\omega_0$). The $Q$ factor of the receiver $Q_3$ is mainly controlled by the load. Indeed, to deliver a meaningful output power, $Q_3\approx Q_w$  and is relatively small, where $Q_w$ is the load \emph{Q} factor. When the system is at resonance ($\omega_\textnormal{in}=\omega_0$), $\tilde{a}_2/\tilde{a}_1$ can be small provided that $\kappa Q_3\gg 1$ and $Q_2\geq Q_3$; which can be achieved if $\kappa$ is made large enough, as in the coupling of a DR with an enclosure.  The magnitude of $\tilde{a}_2/\tilde{a}_1$ then approaches $1/\kappa Q_3$, a small value. It is worth to note that the condition $Q_2\geq Q_3$ can be relaxed for reasonably large values of $\kappa Q_3$  (i.e, $Q_2\geq Q_3$ is a sufficient, but not a necessary condition).

When $\kappa Q_3\gg 1$ and $Q_2\geq Q_3$, the ratio of the fields in DR3 and DR1 ($\tilde{a}_3/\tilde{a}_1$) approaches -1, which is similar to (\ref{eq:Eigenvector}). Hence, for significantly large values of $\kappa$, the non-bonding mode can be excited with high fidelity.

It is worth to mention that there will always be reminiscent contribution from the other coupled modes. To assure that the non-bonding mode is only excited, the system must be prepared in a state in which the fields in DR1 and DR3 are $180^\circ$ out of phase. This is possible if the proper excitation is applied to both DR1 and  the receiver resonator DR3, which defies the purpose of a WPT scheme. It was also shown that by meticulously changing $\kappa$ with time, the system can be made to stay in the non-bonding mode.\cite{Hamam2009}
\subsubsection*{Transfer Efficiency}

In WPT systems, one is mainly interested in the transfer efficiency from the source to the load. The transfer efficiency $\eta\equiv\eta(\omega_0)$ can be written as
\begin{align}
\label{eq:efficiency}
\eta & =\frac{\sigma_w|\tilde{a}_3|^2}{\left(\sigma_w+\sigma_0\right)|\tilde{a}_3|^2+\sigma_0|\tilde{a}_1|^2+\sigma_2|\tilde{a}_2|^2}\\
& = \frac{\sigma_w|\tilde{a}_3(\omega_0)/\tilde{a}_1(\omega_0)|^2}{(\sigma_w+\sigma_0)|\tilde{a}_3(\omega_0)/\tilde{a}_1(\omega_0)|^2+\sigma_0+\sigma_2|\tilde{a}_2(\omega_0)/\tilde{a}_1(\omega_0)|^2}
\end{align}
where $\sigma_w$ models the load absorption. In general, CMT can be used to find expressions for $\eta$.
From (\ref{eq:a2_a1}) and assuming that $\zeta=1/\kappa^2 Q_2Q_3\ll 1$
\begin{equation*}
\frac{\tilde{a}_2}{\tilde{a}_1}\approx \frac{-i}{\kappa Q_3}\left(1-\zeta\right).
\end{equation*}
Similarly,
\begin{equation*}
\frac{\tilde{a}_3}{\tilde{a}_1}\approx -\left(1-\zeta\right).
\end{equation*}
Noting that $Q=\omega/2\sigma$, (\ref{eq:efficiency}) can be written in terms of the $Q$ factors as
\begin{equation*}
\eta=\frac{Q_0\left(1-\zeta\right)^2}{Q_0\left(1-\zeta\right)^2+Q_w+Q_0\zeta\left(1-\zeta\right)^2},
\end{equation*}
where $Q_3\approx Q_w$ was assumed. Noting that $\textnormal{F.O.M}\equiv \kappa^2Q_0Q_2$. The above equation can be written in a compact form as
\begin{equation}
\label{eq:compactEfficiency}
\eta=\frac{\zeta\left(1-\zeta\right)^2 \textnormal{F.O.M}}{\zeta\left(1-\zeta\right)^2 \left(1+\zeta\right)\textnormal{F.O.M}+1}.
\end{equation}
The maximum efficiency $\eta^\textnormal{max}$ can be determined by equating the derivative of the above equation to zero (i.e, $\partial\eta/\partial\zeta|_{Q_w=Q_w^\textnormal{max}}=0$), which is equivalent to finding the roots of a fifth order polynomial. However since $\zeta\ll 1$, one can ignore orders higher than two and solve the quadratic equation
\begin{equation}
\zeta^2+3\textnormal{F.O.M}^{-1}\zeta-\textnormal{F.O.M}^{-1}=0
\end{equation}
to estimate that for large values of $\textnormal{F.O.M}$ (usually the case), the above polynomial has a root which approaches $\zeta_0\equiv 1/\sqrt{\textnormal{F.O.M}}$. Hence,
\begin{equation}
\label{eq:Qmax}
Q_w^\textnormal{max}\approx\frac{1}{\kappa}\sqrt{\frac{Q_0}{Q_2}}
\end{equation}
The maximum efficiency $\eta^\textnormal{max}$ can be found by substituting $\zeta_0$ back in (\ref{eq:compactEfficiency})
\begin{equation}
\label{eq:Etamax}
\eta^\textnormal{max}\approx1-\frac{2}{\sqrt{\textnormal{F.O.M}}}.
\end{equation}

From (\ref{eq:Qmax}) and (\ref{eq:Etamax}), it is clear that increasing $\kappa$ or $Q_2$ is desirable to get more useful power out of the system (decreasing $Q_w^\textnormal{max}$). In this case, the maximum efficiency increases as well. The  Figure Of Merit quantifies the performance of the coupled system. However for the three centre system, it depends on $Q_2$ rather than on the load $Q_w$, which is different from the case of two inductively coupled resonators.\cite{MIT_Theory} This observation emphasizes the role of the enclosure as a mediator.

Inside the $\kappa Q_3\gg 1$ regime, the non-bonding mode is excited with high fidelity, which allows us to simplify $\eta$ given by (\ref{eq:efficiency}) to
\begin{equation}
\label{eq:efficiencyRes}
\eta=\frac{\sigma_w}{\sigma_w+2\sigma_0}.
\end{equation}

For $\sigma_w\gg\sigma_0$, $\eta$ asymptotically approaches unity. It worth to note that when $\sigma_w=2\sigma_0$, the efficiency is 0.5; this physically means that energy absorbed by the load is twice that absorbed by each DR, with no energy absorbed by the enclosure since its fields are not excited.
\subsection{Circuit Modelling}
\begin{figure}[!h]
\begin{circuitikz}[scale=0.18]
\ctikzset {bipoles/length=0.54cm}
\draw
 (0,0) to [sV, l^=$V_s$] (0,4)
 to [R, l^=$R_s$] (0,8) to (0,9) to(1,9)
 to [R, l^=$R_0$] (5,9)
 to [L, l^=$L_0$] (9,9)
 to [C, l_=$C_0$] (9,0) 
 to (0,0);
 \draw [fill=yellow,yellow, opacity=0.5] (14,-1) rectangle (25,10);
 \draw
 (15,0) to [C, l^=$2C_r$] (15,9) to (16,9)
 to [R, l^=$R_r$] (20,9)
 to [L, l^=$L_r$] (24,9)
 to [C, l_=$2C_r$] (24,0) 
 to (15,0);

 \draw
  (30,0) to [C,l_=$C_0$] (30,9) to (31,9)
 to [R, l^=$R_0$] (35,9)
 to [L, l^=$L_0$] (39,9)
 to [R, l^=$R_L$] (39,0)
 (30,0) to (39,0);
 \draw [fill=white] (9,9.5)node(a){} circle (0pt);
\draw [fill=white] (24,9.5)node(b){} circle (0pt);
  \draw [<->,>=stealth] (a)  to [bend left] node[pos=0.6] {} ++(6,0);
    \draw [<->,>=stealth] (b)  to [bend left] node[pos=-0.6] {} ++(6,0);
    (19.5,0) node{(a)}
    \draw (12,9) node{$C_m$};
    \draw (27,9) node{$C_m$};
 \end{circuitikz}
\caption{Circuit Model of three coupled resonators, highlighting the relay resonator. The electric coupling is modelled by the mutual capacitance $C_m$.}
\label{fig:CircuitModel}

\end{figure}
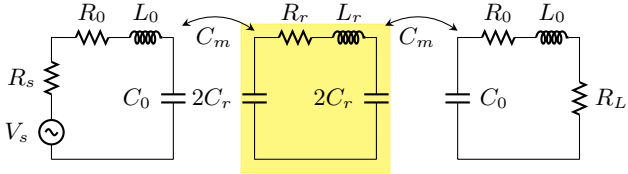

\begin{figure}[!h]
  \centering
   \includegraphics[width=3.5in]{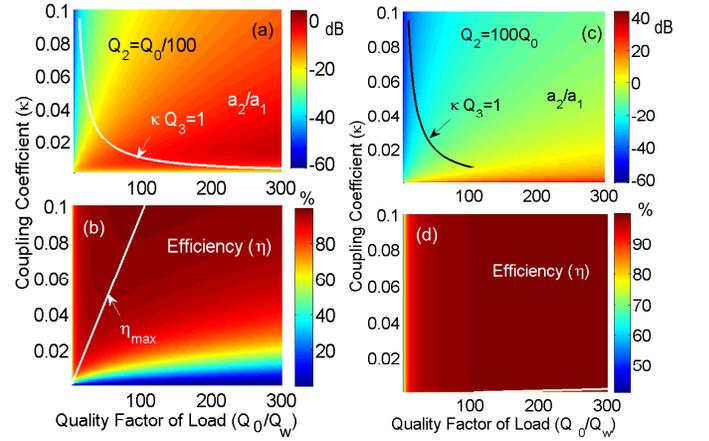}
   \caption{(Top) The ratio in decibel of the relay current $i_R$ to the source current $i_S$ as a function of the coupling coefficient $\kappa$ and the load $Q_w$. The white curve represents the $\kappa Q_3=1$ condition. (Bottom) The transfer efficiency $\eta$. The white curve depicts the locus of the $\eta^\textnormal{max}$, as given by (\ref{eq:Qmax}); which is the maximum efficiency for each $\kappa$ value. }
  \label{fig:Colorful}
\end{figure}

In this subsection, the three centres system is modelled as three coupled LC oscillators. Fig.\ref{fig:CircuitModel} shows the schematic of the circuit model. Because of symmetry, the relay capacitance $C_r$ is split into two series capacitance of $2C_r$ each. Solving for the mesh current, the efficiency and the $i_R/i_S$ can be determined for different loads $R_L$ and coupling capacitance $C_M$. For this particular circuit, $\kappa$ can be determined from the free running condition (letting $V_s=0$, see Appendix B),
\begin{equation}
\kappa=\frac{C_m}{2\sqrt{C_0C_r}}.
\end{equation}
Once the currents are known, the energy density in each resonator ($\propto |i|^2$) can be obtained. The ratio of energy in the mediator (relay) to that in the input DR as a function of the coupling coefficient ($\kappa=C_m/2\sqrt{C_0C_r}$) and Quality factor of the load $Q_w=\omega L_0/R_L$ is plotted in Fig. \ref{fig:Colorful} for different values of $\kappa$ and $Q_w$. The \emph{slight} shift in the frequency of the non-bonding mode due to the strong interaction is automatically taken into account. From the Fig., it is clear that as $\kappa$, and $Q_w$ increase (moving to the top left corner of Fig. \ref{fig:Colorful}(a)), the energy stored in the relay  diminishes which corraborates with the CMT prediction. When $Q_2\gg Q_w$, which is the case depicted in Fig. \ref{fig:Colorful}(c), the energy in the relay  is diminished further. For a fixed $Q_w$ value, $\eta$ increases with the increase of $\kappa$. This is because the system is pushed deeper into the $\kappa Q_3\gg 1$ regime and the effect of the non-bonding mode is more profound. The white line in Fig. \ref{fig:Colorful}(b) identifies the loci where maximum efficiency is attained. Agreeing with (\ref{eq:Qmax}), higher values of $\kappa$ correspond to more absorbed load power (lower $Q_w^\textnormal{max}$). Moreover, the value of $\eta^\textnormal{max}$ increases with $\kappa$; a manifestation of (\ref{eq:Etamax}). Even lower $Q_w^\textnormal{max}$ values can be achieved by increasing $Q_2$. In this case, $\eta^\textnormal{max}$ is pushed to the right. For the situation depicted in Fig. \ref{fig:Colorful} (d),  $Q_2$ increases $10^4$ folds, which according to (\ref{eq:Qmax}) means that $Q_w^\textnormal{max}$ decreases by two orders of magnitude. Thus, the $\eta^\textnormal{max}$ locus is pushed to the far right outside of the domain of interest and $\eta^\textnormal{max}$ approaches unity.

\begin{figure}[!h]
  \centering
   \includegraphics[width=2.75in]{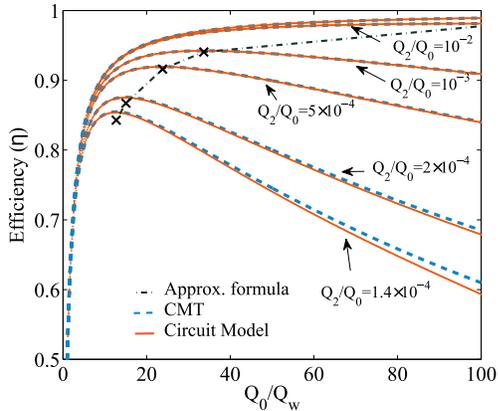}
   \caption{The efficiency $\eta$ curves calculated for different values of $Q_2$ using the circuit model and the approximate formula (\ref{eq:compactEfficiency}). The coupling coefficient $\kappa$ is set to 0.1.}
\label{fig:EfficiencyFormulae}
\end{figure}

Fig. \ref{fig:EfficiencyFormulae} shows how the efficiency changes as a function of the load $Q$. It is clear that the efficiency attains  maxima, especially for lower $Q_2$ values. The positions and values of these maxima can be accurately estimated using (\ref{eq:Qmax}) and (\ref{eq:Etamax}) as shown in the Fig. To calculate $\eta$ using the circuit model, the shift of the resonant frequency of the non-bonding mode due to coupling was determined from (\ref{eq:exact_omega}). For $\kappa=0.1$, the corresponding frequency shift is $\sim 1\%$.

\subsection{Transient Analysis}
\begin{figure}[!h]
  \centering
   \includegraphics[width=2.6in]{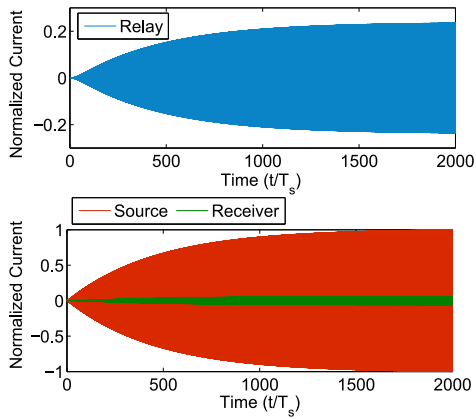}
   \caption{Transient Response of three coupled resonators for $\kappa=0.0024$, $Q_2/Q_0$= 1/100. (Top) The current in the Relay (Mediator) resonator. (Bottom) Current in Source and Receiver Resonators. All current values are normalized to the peak of the source steady state value. $T_s$ is the period of the excitation and is equal to $2\pi/\omega_\textnormal{in}$. The values of $Q_0$, $Q_2$ and $Q_w$ are $\approx 10,000$, $\approx100$  and $\approx 105$, respectively. }
  \label{fig:TransientLowK}
\end{figure}

\begin{figure}[!h]
  \centering
   \includegraphics[width=2.6in]{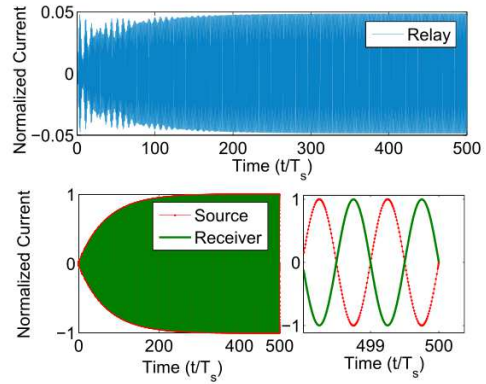}
   \caption{Transient Response of three coupled resonators for $\kappa=0.1897$, $Q_2/Q_0$= 1/100. (Top) The current in the Relay (Mediator) resonator. (Bottom) Current in Source and Receiver Resonators overlap. All current values are normalized to the peak of the source steady state value. $T_s$ is the period of the excitation and is equal to $2\pi/\omega_\textnormal{in}$. The values of $Q_0$, $Q_2$ and $Q_w$ are $\approx 10,000$, $\approx100$  and $\approx 105$, respectively.}
  \label{fig:TransienthighK}
\end{figure}

So far, the theoretical analysis, either using CMT or circuit modelling was concerned with the steady state response. It says nothing about how the fields values inside the receiver evolve from initial time, until the response is dominated by the non-bonding mode. To examine the evolution of the fields inside the resonators from the moment the input source is turned on to steady state, one must resort to transient analysis. For the three coupled resonators, we use state space modelling (See Appendix B), where the system is represented by a set of first order coupled ordinary differential equations (ODEs). The system is allowed to evolve from $t=0$, when the power is turned on till steady state is reached. Figs. \ref{fig:TransientLowK} and \ref{fig:TransienthighK} show the time evolution of the currents for two different values of $\kappa$. In the first case (Fig. \ref{fig:TransientLowK}), $\kappa$ is very small ($\approx 0.0024$) and $\kappa Q_3<1$, indicating that the non-bonding mode is not dominant. From Fig.\ref{fig:TransientLowK}, it is clear that the output current is small, when compared to the current in the mediator. This means that the transfer of energy is inefficient.

The $\kappa^2 Q_2 Q_3\gg 1$ condition is violated when $\kappa$ is small. (For the case presented in Fig. \ref{fig:TransientLowK}, $\kappa^2 Q_2 Q_3\approx0.06$). According to  (\ref{eq:a2_a1}), the magnitude of the normalized mediator current reduces to $\approx \kappa Q_2$, which is equal to 0.25 and agrees with the magnitude at steady state determined by the transient analysis as Fig. \ref{fig:TransientLowK} shows.

However when $\kappa$ increases (Fig. \ref{fig:TransienthighK}), energy quickly builds up in the receiver and the steady state response is dominated by the non-bonding mode. The fields in both the source and receiver are approximately equal in value and $180^\circ$ out of phase. Moreover, the fields in the mediator are significantly reduced (For the parameters used in Fig. \ref{fig:TransienthighK}, the steady state value of the mediator current is $1/\kappa Q_3=0.05$ less than the value in either the source or receiver, which agrees with the steady state value calculated from the transient analysis). 

At the transient epoch, the spectrum of the source as it turns on is essentially wide band and it excites other coupled and spurious modes. These modes are essentially responsible for building up energy inside the receiver. When the frequency values of the modes are sufficiently different from the non-bonding mode, their effect diminishes and the steady state response is dominated by the non-bonding mode.
\section{Results and Discussion}
In this section two systems are studied using full wave finite element computation. In the first system, power is transferred from one point to another inside an enclosure using the interaction between it and two DRs. Hence, the system is denoted by 2DR/CV. The second system exploits the interaction between an SCR mode and DRs to transfer power over an air gap with minimal fringing fields. In this particular case the structure is open.

For both systems, the analysis carried out in the previous section is still applicable. With no loss of generality, the enclosure $\textnormal{TE}_\textnormal{011}$ mode and DRs $\textnormal{TE}_{01\delta}$ modes are used. The resonant frequencies are determined using HFSS\textsuperscript{\textregistered} Eigenmode solver (Ansys Corporation, Pittsburgh, PA, USA). The system dimensions are tuned to assure that all resonators have the same resonant frequencies $\omega_0$. The driven mode solver is then invoked to determine the transfer efficiency $\eta$. Based on the argument of the previous section, the values of the output port characteristic impedance $Z_{02}$ at which $\eta=0.5$ is numerically found ($Z_{02}^*$). In this case the 0.5 efficiency corresponds to $\sigma_w^*=2\sigma_0$; hence the $\sigma_w$ can be controlled by changing $Z_{02}$ as
\begin{equation}
\sigma_w=2\frac{Z_{02}}{Z_{02}^*}\sigma_0.
\end{equation}
The transfer efficiency is calculated from
\begin{equation}
\eta=\frac{|S_{21}|^2}{1-|S_{11}^2|}.
\end{equation}
\subsection{2DR/CV Interaction}
 For this system, the enclosure is a cylindrical cavity. It was previously shown that such interaction can be strong and $\kappa$ can be substantially large.\cite{Paper5}  Transferring power inside an enclosure, by exploiting the enclosure modes gained recent interest due to the ability of transferring power to any point in 3D space and long transfer distance.\cite{SampleAPL, Mei2016_1, Mei2016_2, Wu2016, Lin2017} In the following discussion, we will discuss point to point transfer efficiency, where the enclosure modes merely act as mediators (i.e, exploiting the non-bonding mode). The main advantage of using the non-bonding mode as a vehicle for power transfer is that the energy is localized within the DRs, with minimal penetration in the 3D space.
\begin{figure}[!h]
  \centering
   \includegraphics[width=3.0in]{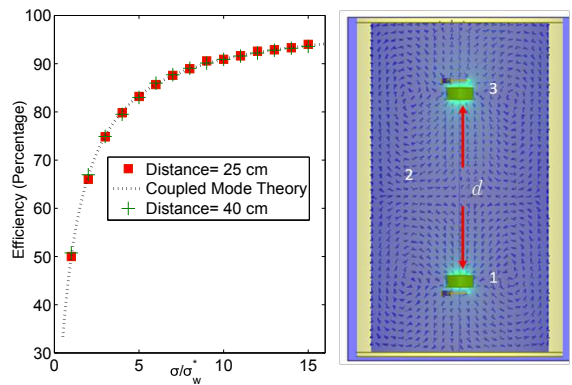}
   \caption{(Left) Efficiency vs. relative load $\sigma_w/\sigma_w^*$ for $d=$ 25 cm ($\approx0.8\lambda$) and 40 cm ($\approx1.3\lambda$). (Right) The system configuration. The DRs are identical; $\epsilon_r=29.2$, height = 26.5 mm, diameter = 30 mm, $f_{TE_{01\delta}}\approx 960 \textnormal{ MHz}$ and $\tan\delta= 10^{-4}$. The enclosure is a cylindrical cavity made of Cu; height = 75 cm, diameter = 38.4 cm, $f_{TE_{011}}\approx 960 \textnormal{ MHz}$.}
\label{fig:DR2CV}   
\end{figure}

Figure \ref{fig:DR2CV} shows the computed efficiency, which agrees with (\ref{eq:efficiencyRes}). It is important to emphasize that the efficiency does not strongly depend on the transfer distance; also the transfer distance is significant when compared to the wavelength. To show the effect of the cavity as a mediator, $\eta$ is computed for the two DRs only, when placed 25 cm apart and the cavity is removed. The efficiency is considerably small ($\approx 0.025\%$), emphasizing the role of the cavity as a mediator. Similarly, $\eta$ was found to be 5\% when the cavity is only present and the two DRs are removed, still considerably lower than when the both the cavity and DRs are present ($\approx 94\%$).

It is essential to validate that the system depicted in Fig. \ref{fig:DR2CV} does indeed operate in the $\kappa Q_3\gg 1$ regime. First, $\kappa_0$, the value of $\kappa$  when a DR is placed at the cavity center was determined to be $\approx 0.097$. Because the electric field of the $\textnormal{TE}_{011}$ mode has a $\cos(\pi z/H)$ dependency, $\kappa$ at a distance $d$ from the cavity center can be approximated by
\begin{equation}
\kappa(d)=\kappa_0\cos{\frac{\pi d}{2H}},
\end{equation}
where $H=$ 75 cm is the cavity height. For $d= 40 \textnormal{ cm}$, $\kappa\approx 0.065$. Therefore, $Q_3$ should be significantly larger than 15.4 ($1/\kappa$) to guarantee that the system is in the $\kappa Q_3\gg 1$ regime. Note that $Q_0=1/\tan\delta =10,000$ and that $Q_w$ at 0.5 efficiency ($Q_w^*$) is $Q_0/2$. The lowest possible $Q_3$ occurs when $Q_w$ is the smallest; this happens at $\sigma_w/\sigma_w^*=15$. Hence the lowest $Q_3$ is approximately $333$, substantially larger than 15.4, which guarantees that the system operates in the $\kappa Q_3\gg 1$ regime over the entire simulated loads (the abscissa of Fig. \ref{fig:DR2CV}). It is worth to mention that $Q_2$ is large because the cavity was made of copper. However for SCRs, $Q_2$ can be considerably smaller due to radiation losses.

\subsection{2DR/SCR Interaction}
In this configuration, the enclosure is a SCR. Historically, this structure was introduced as a component to measure the dielectric properties of a specimen sandwiched  between the two halves.\cite{Janezic99} The SCR mode of interest here is a modified $\textnormal{TE}_\textnormal{011}$. As long as the gap between the two halves is small, the fields are confined within the structure, with minimal leakage. This can be understood if one considers the two flanges to act as a parallel plates waveguide. For small gaps, the waveguide is below cut-off and the fields are confined. In general, the shorter the flanges, the more the leakage; and the lower the \emph{Q}.

\begin{figure}[!h]
  \centering
   \includegraphics[width=2.65in]{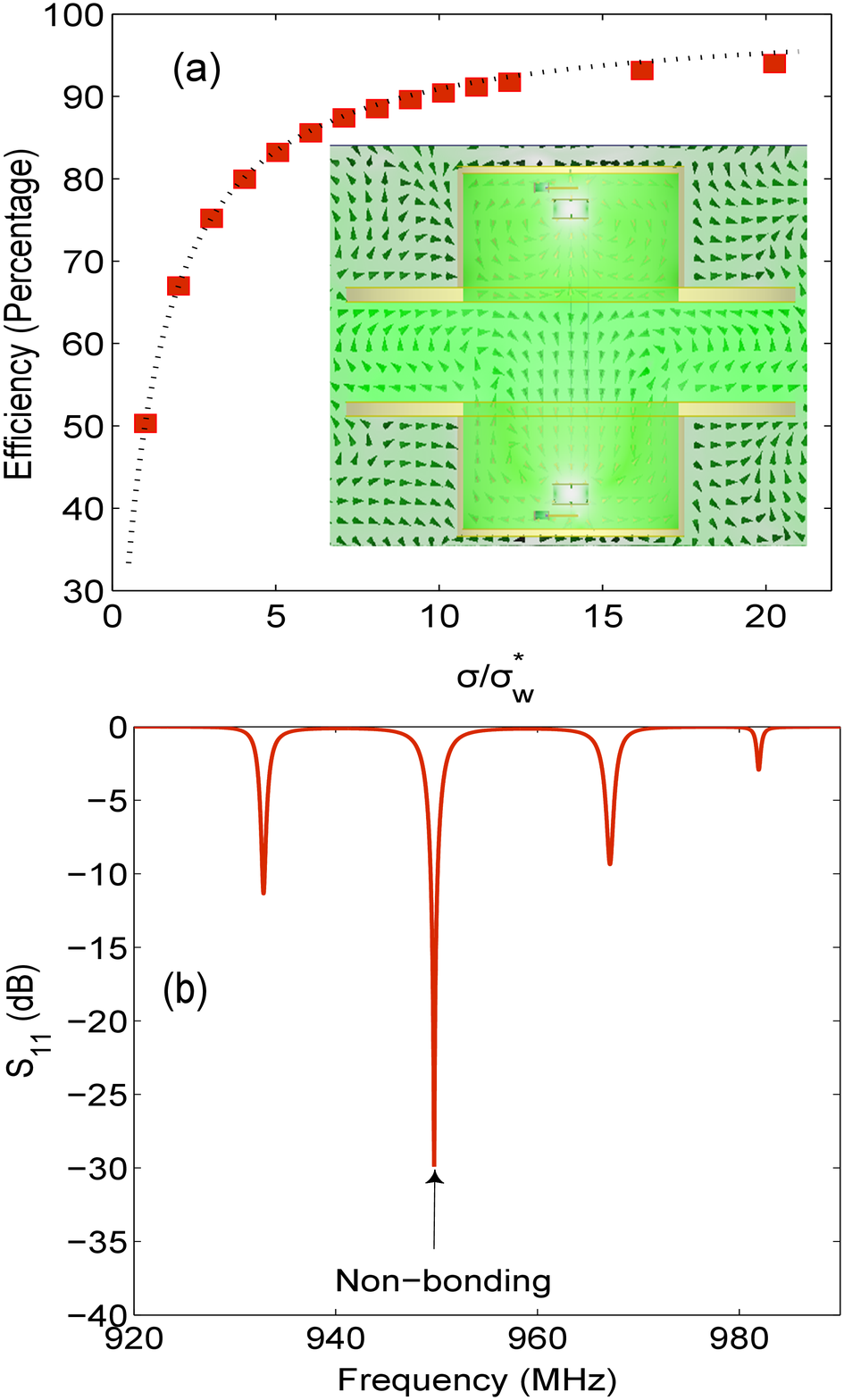}
   \caption{(a) The efficiency of a 2DR/SCR as a function of the relative load $\sigma_w/\sigma_w^*$. Inset: The geometry and the magnetic field distribution of the non-bonding mode. The SCR has the following dimensions: height = 50 cm, diameter= 37 cm, Flange extension: 20 cm and the gap = 14 cm. Each DR has the following parameters: $\epsilon_r=29.2$, height = 27 mm, diameter= 31 mm; and its $\tan\delta=10^{-4}$. The resonant frequency of the DRs and SCR is around 950 MHz. (b) The reflection coefficient $S_{11}$ as a function of the operating frequency. The arrow identifies the non-bonding mode.}
\label{fig:DR2SCR}
\end{figure}
\begin{figure}[!h]
  \centering
   \includegraphics[width=2.5in]{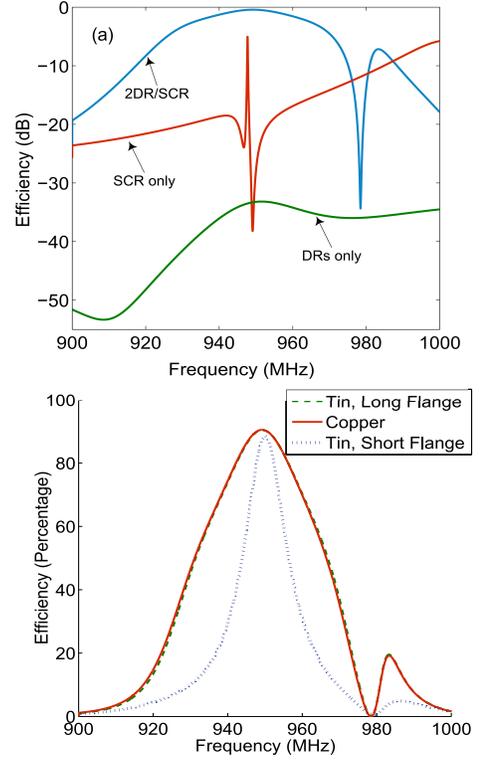}
   \caption{(Top) The computed efficiency, expressed in dB, as a function of the operating frequency. 0 dB is equivalent to unity. (Bottom) Efficiency as a function of frequency for three different configurations according to the SCR material and flange extension (20 and 5 cm).}
\label{fig:DR2SCR_Tin}
\end{figure}

Similar to the previous subsection, the efficiency was calculated for different loads and different input frequency, as depicted in Fig. \ref{fig:DR2SCR}. Again, there is a good agreement between the computed and predicted efficiency as given by (\ref{eq:efficiencyRes}) over the relative load values $\sigma/\sigma^*=$ 1 to 20. Noting that $\kappa_0\approx 0.1$ and that DRs are separated from the SCR center by 18.7 cm, it can be shown, identical to the analysis carried out in the previous section, that the system does indeed operate in the $\kappa Q_3\gg 1$ regime. In Fig. \ref{fig:DR2SCR}(b), the other two coupled modes appear as dips surrounding the non-bonding mode. The small dip at $\approx$ 980 MHz is a higher SCR mode. The efficiency versus frequency plotted in Fig. \ref{fig:DR2SCR_Tin}(a) illustrates that the 2DR/SCR structure has a remarkable efficiency in a wide band determined by the coupled modes, where it attains its maximum value at the non-bonding frequency. This is to be contrasted to the low efficiency value due to the cavity mode only, which has a narrow band  due to the excitation of the $\textnormal{TE}_\textnormal{011}$ mode only. Moreover, the efficiency of the two DRs in the absence of the SCR is much lower; since they are practically uncoupled.

The non-bonding mode has the intriguing property of zero enclosure excitation. Such property allows us to down grade the performance of the enclosure (i.e, reducing its \emph{Q}). In Fig \ref{fig:DR2SCR_Tin}(b), the SCR material was changed from copper to tin which has a significantly lower conductivity. In this case, as expected, the efficiency did not  significantly change. Moreover, the flange length was reduced from 20 to 5 cm which means that more power can leak from the SCR mode and hence decrease its \emph{Q}. Again, the efficiency at the non-bonding frequency did not significantly change, unlike its value at other frequencies.

\begin{figure}[!h]
   \centering
   \includegraphics[width=2.5in]{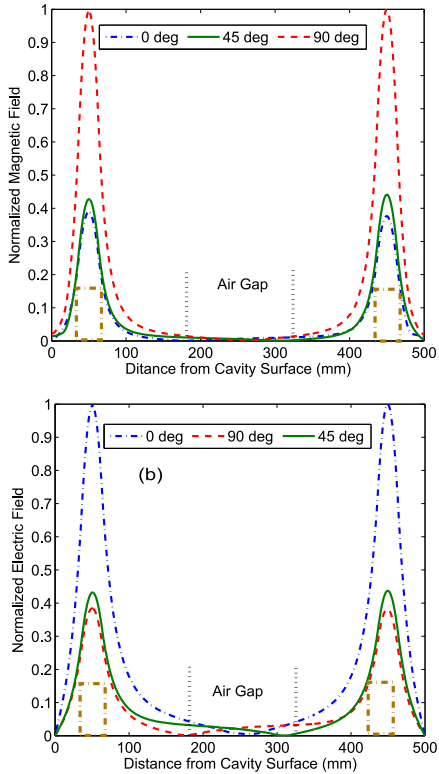}
   \caption{Normalized fields at different phases. (Top) Normalized Magnetic Field along the SCR axis. (Bottom) Normalized axial Electric Field at the DRs radius. The DRs are denoted by the dashed-dotted rectangles below the peaks.}
\label{fig:EH_DR2SCR}   
\end{figure}

Under the $\kappa Q_3\gg 1$ regime the fields of the enclosure mode are minimal. On the other hand the fields of the DR modes are excited and are confined within the dielectric material with an evanescent profile outside it.\cite{Pozar05} The DRs in Fig. \ref{fig:DR2SCR} (a) are placed deep inside the cavity to assure that their fringing fields are minimal in the gap. Fig. \ref{fig:EH_DR2SCR} depicts the computed axial electric and magnetic fields. The Fig shows that the fields amplitudes decrease at the gap; this can be beneficial in biomedical and high power applications where the electric and magnetic fields can cause safety hazards.

\begin{figure}[!h]
  \centering
   \includegraphics[width=2.65in]{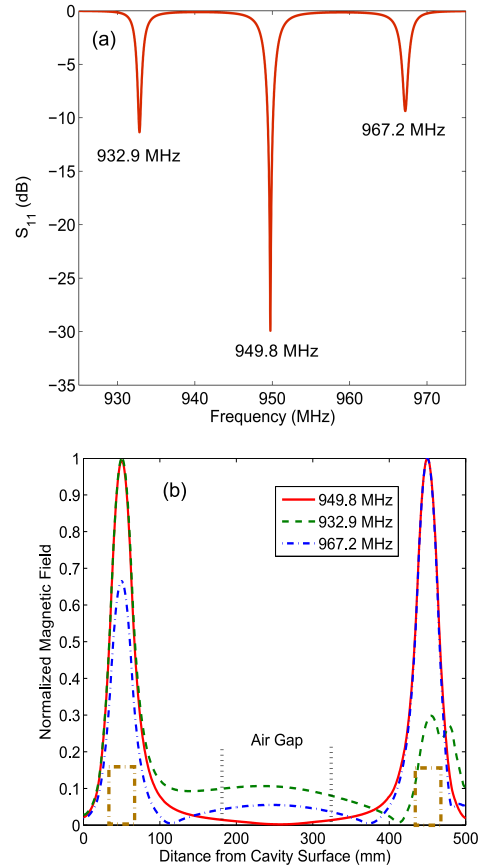}
      \caption{(a) Reflection Coefficient $S_{11}$ showing the three coupled modes. (b) The axial normalized magnetic field of the three modes. The DRs are denoted by the dashed-dotted rectangles below the peaks.}
\label{fig:ThreeModes}    
\end{figure}

To further clarify the unique properties of the non-bonding fields when compared with the other two coupled modes, Fig. \ref{fig:ThreeModes} shows the normalized magnetic field of the three modes. From the Figure it is clear that, unlike the non-bonding mode, the SCR mode manifests itself in the lower and higher modes.

The dimensions and properties of the 2DR/SCR structure can be optimized for better efficiency, transfer distance and compactness. This can be achieved, for example, by optimizing the DR dimensions and dielectric constant. Although the transfer distance ($\approx\lambda/2$) is large when compared to the usual sub-wavelength regime of resonant inductive  coupling devices, this usually comes with the price of large SCR dimensions. The SCR dimensions can be reduced by shortening the flange extrusion as was shown in Fig. \ref{fig:DR2SCR_Tin} (b). It can be further reduced via the embedding of a suitable dielectric material. The gap width can also be increased by raising its cut-off frequency; may be by engineering corrugations on its surface which acts as an effective epsilon near zero surface in the vicinity of the resonance frequency. This has the effcect of increasing the cut-off frequency of the parallel plates waveguide formed by the upper and lower flanges. In some applications, however, the 2DR/SCR structure can take advantage of the metals already present and may be unavoidable.
 
\section{Conclusion}
The non-bonding mode, resulting from the interaction of two dielectric resonators with an enclosure, induces an electromagnetic induced transparency like process capable of efficiently transferring power over a distance comparable to $\lambda$. Unlike resonant inductive coupling, the coupling is mainly electric. Coupled Mode Theory and  Lumped Circuit models were used as the analytical tools to determine the system parameters such as the optimal load $Q_w^\textnormal{max}$ and the corresponding maximum efficiency $\eta^\textnormal{max}$. The strong coupling between the dielectrics and the enclosure makes it possible to excite the non-bonding with high fidelity. The efficiency of the system is predicted to be high and asymptotically approach unity. It does not depend on the transfer distance even when the distance is of the order of $\lambda$. Thus, suggesting that the analysis developed here can be used to design WPT systems that can efficiently power autonomous systems placed in a controlled environment. During steady state the enclosure mode has insignificant contribution to the total fields. Hence, it is possible to use low profile enclosures and design systems which have negligible fringing fields. This suggest its use in biomedical or higher power applications, where the exposure to the electromagnetic fields is of great concern. During the transient epoch, briefly after the input excitation is turned on, the energy in the receiver builds up via all coupled and spurious modes. However since they are off-resonance, the non-bonding mode only sustains in the steady state response.
\section*{APPENDIX A: Forced Energy CMT}
Here we present an outline of how the coupled mode equation (\ref{eq:forcedECMT}) was obtained. For an elaborate discussion, please refer to Ref. \cite{URSI2016}. The $\mathbf{E}$ and $\mathbf{H}$ fields of the coupled modes are expanded in terms of the uncoupled ones ($\mathbf{E}_k$ and $\mathbf{H}_k$), where $a_k\left(t\right)$ and $b_k\left(t\right)$ are time dependent to take into account the effect of the time dependent impressed source ($\mathbf{J}_{\textnormal{imp}}$). Accordingly,

\begin{equation}
\label{eq:EHfield}
\mathbf{E}=\sum_{k=1}^N a_k\left(t\right)\mathbf{E}_k\textnormal{  and   } \mathbf{H}=\sum_{k=1}^N b_k\left(t\right)\mathbf{H}_k,
\end{equation} 
where $N$ is the number of interacting modes ($N=3$ for the three centres system). As was carried out in Ref. \cite{Paper2}, the coupled mode equations can be determined by expanding $\nabla\cdot\left(\mathbf{E}_k^*\times\mathbf{H}\right)$ and $\nabla\cdot\left(\mathbf{E}\times\mathbf{H}_k^*\right)$, using Maxwell's equations and (\ref{eq:EHfield}). The time dependency of $a_k\left(t\right)$ and $b_k\left(t\right)$, and the presence of $\mathbf{J}_\textnormal{imp}$ generalize the coupled equations to

 \begin{equation}
\label{eq:CEqn1}
\A\dot{\bm{a}}+\left(\M+\F-i\Omega\B\right)\bm{b}=\J
\end{equation}

\begin{equation}
\label{eq:CEqn2}
\G\dot{\bm{b}}+\left(\M^\dag-i\Omega\D\right)\bm{a}=0,
\end{equation}
where $\A$, $\M$, $\F$, $\B$, $\G$ and $D$ are $N\times N$ matrices that are functions of the overlap integrals and are defined in\cite{Paper2}. $\Omega$ stores the uncoupled complex frequencies and $\J$ is an $N\times 1$ column vector which includes the forcing terms due to the interaction of the modes with $\mathbf{J}_\textnormal{imp}$. Its $k^\textnormal{th}$ row is given by 

\begin{equation*}
\label{eq:Jimp}
\J_k=-\int_V\mathbf{J}_\textnormal{imp}\cdot\mathbf{E}_k^*~dv.
\end{equation*}
Eqs. (\ref{eq:CEqn1}) and (\ref{eq:CEqn2}) are coupled differential equations in the fields amplitudes. Under the assumption that radiation losses are small, one finds that
\begin{equation}
\label{eq:ECMT_forced_simplified}
\ddot{\bm{a}}+\K\bm{a}=\A^{-1}\dot{\J},
\end{equation} 
where $\K=\A^{-1}\D^\dag\Omega\G^{-1}\Omega\D$.\cite{Paper2} Taking the Laplace transform of (\ref{eq:ECMT_forced_simplified}), the response can be expressed in the $s-$domain as
 
\begin{equation}
\label{eq:ECMTsd}
\mathcal{L}\{\bm{a}\}=\bm{\Phi}\left(s\right)\left[s\dot{\bm{a}}\left(0\right)+\bm{a}\left(0\right)\right]+s\bm{\Phi}\left(s\right)\mathcal{L}\{\J\},
\end{equation} 
where $\bm{\Phi}\left(s\right)\equiv\left(s^2\bm{I}+\K\right)^{-1}$ is the $N\times N$ matrix transfer function. When the system is excited by sinusoidal inputs of frequency $\omega$, the steady state response is determined by the last term in the R.H.S of (\ref{eq:ECMTsd}) as 

\begin{equation}
\label{eq:ECMTw}
\tilde{\bm{a}}\left(\omega\right)=i\omega\bm{\Phi}\left(\omega\right)\mathcal{L}\{\J\}|_{s=i\omega},
\end{equation}
where $s$ is replaced by $i\omega$.

\section*{APPENDIX B: Analysis of Circuit Model}
\begin{figure}[!h]
\begin{circuitikz}[scale=0.18]
\ctikzset {bipoles/length=0.6cm}
 \draw
   (0,-14) to [sV, l^=$V_s$] (0,-10) 
   to [R, l^=$R_s$] (0,-6) to (0,-5) to (1,-5)
   to [R, l_=$R_0$,i=$i_S$] (5,-5) to [L, l_=$L_0$] (9,-5)
   to [C, l_=$C_0'$] (9,-14)
   (9,-5) to (10,-5)
   (10,-5) to [C, l_=$C_m$, i=$i_1$] (14,-5)
   to [C, l^=$2C_r'$] (14,-14)
   (14,-5) to (15,-5)
    (15,-5) to [L, l_=$L_r$, i=$i_R$] (19,-5) 
   to [R, l_=$R_r$] (23,-5) to (24,-5)
   to [C, l^=$2C_r'$] (24,-14) 
   (24,-5) to (26,-5) to [C, l_=$C_m$, i=$i_2$](30,-5) to (31,-5)
    to [C, l^=$C_0'$] (31,-14)
   (31,-5) to [L, l_=$L_0$] (35,-5)
   to [R, l_=$R_0$,i=$i_L$] (39,-5) to (41,-5)
   to [R, l^=$R_L$](41, -14)
   (0,-14) to (41,-14)
   (10,-4) node{$v_1$}
   (15,-4) node{$v_2$}
   (24,-4) node{$v_3$}
   (31,-4) node{$v_4$};

 \end{circuitikz}
\caption{Equivalent circuit of the coupled resonators in Fig. \ref{fig:CircuitModel}. The coupling capacitance $C_m$ is replaced by an equivalent $\bm{\pi}$ network.}
\label{fig:CircuitModelAppendix1}
\end{figure}
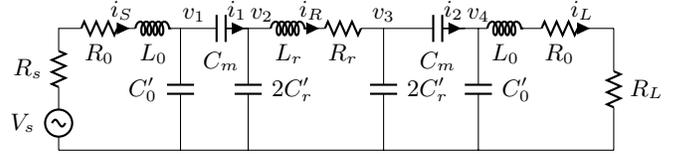

Fig. \ref{fig:CircuitModelAppendix1} shows the equivalent circuit of the three coupled LC circuits (Fig. \ref{fig:CircuitModel}).

The currents in each loop can be determined after solving the mesh equations

\small
\begin{align*}
&\left(R_s+R_0+j\omega L_0-\frac{j}{\omega C_0'}\right)i_S+\frac{j}{\omega C_0'}i_1=V_s,\\
&\frac{j}{\omega C_0'}i_S+\left(\frac{-j}{\omega C_m}-\frac{j}{2\omega C_r' }-\frac{j}{\omega C_0'}\right)i_1+\frac{j}{2\omega C_r'}i_R=0,\\
&\left(-\frac{j}{2\omega C_r'}+j\omega L_r+R_r -\frac{j}{2\omega C_r'}\right)i_R+\frac{j}{2\omega C_r'}i_1+\frac{j}{2\omega C_r'}i_2=0,\\
&\frac{j}{2\omega C_r'}i_R+\left(-\frac{j}{\omega C_m}-\frac{j}{\omega C_0'}-\frac{j}{2\omega C_r'}\right)i_2+\frac{j}{\omega C_0'}i_L=0, \textnormal{and}\\
&\left(-j\frac{1}{\omega C_0'}+j\omega L_0+R_0+R_L\right)i_L+\frac{j}{\omega C_0'}i_2=0,
\end{align*}
\normalsize
where $C_0'\equiv C_0-C_m$ and $2C_r'\equiv 2C_r-C_m$.

\subsection*{Determine $\kappa$ from the Circuit Model}
Using (\ref{eq:EigenvalueProb}), CMT predicts that the frequencies of the coupled system are given by
\begin{align*}
\omega_1&=\omega_0,\\
\omega_2&=\omega_0\sqrt{1+\sqrt{2}\kappa},\\
\omega_3&=\omega_0\sqrt{1-\sqrt{2}\kappa}.
\end{align*}
\begin{figure}[!h]
\centering
\begin{circuitikz}[scale=0.2]
\ctikzset {bipoles/length=0.6cm}

\draw
 (0,0) to [L, l^=$L_0$] (0,8)
 to (6,8)
 to [C, l_=$C_0'$] (6,0)
 (6,8) to [C, l_=$C_m$] (12,8)
 to [C, l_=$2C_r'$] (12,0)
 (12,8) to [L, l_=$L_r/2$] (18,8);
 \draw [fill=yellow, dotted] (18,10) to (18,-2) (17.5,10) node{T} (17.5,-2) node{T'};
 \draw
 (18,8) to [L, l_=$L_r/2$] (24,8)
 to [C, l_=$2C_r'$] (24,0)
 (24,8) to [C, l_=$C_m$] (30,8)
 to [C, l_=$C_0'$] (30,0)
 (30,8) to(36,8)to [L, l_=$L_0$] (36,0)
 (0,0) to (36,0);

\end{circuitikz}
\caption{Equivalent circuit when the input is zero.}
\label{fig:CircuitModelAppendix2}
\end{figure}
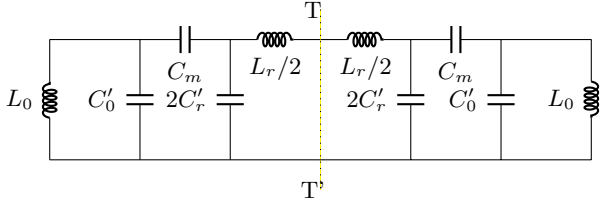
To determine the frequencies using the circuit model, we consider the lossless case. The system is depicted in Fig. \ref{fig:CircuitModelAppendix2}. Because of symmetry around the plane $T-T'$, two cases are considered: (1) $T-T'$ is a Perfect Magnetic wall ($i_R=0$) and (2) $T-T'$ is a Perfect Electric wall ($v=0$). The Perfect Magnetic wall case corresponds to the mode where the relay mode is not excited ($i_R=0$). Accordingly, the equivalent capacitance seen from either sides of $T-T'$ is
\begin{equation*}
C_{eqv}=\frac{2C_mC_r'}{2C_r}+C_0'=C_0-\frac{C_m^2}{2C_r}
\end{equation*}
Therefore, the resonant frequency $\omega_r$ will be
\begin{equation}
\label{eq:exact_omega}
\omega_r=\frac{1}{\sqrt{L_0C_{eqv}}}=\frac{\omega_0}{\sqrt{1-C_m^2/2C_rC_0}},
\end{equation}
indicating that, unlike the approximate CMT analysis, the circuit model predicts a slight blue shift in the frequency of the non-bonding mode.

When the symmetry plane is a Perfect Electric wall, the equivalent circuit is shown in Fig. \ref{fig:CircuitModelAppendix3}. The resonance frequency in this case can be found by calculating the total impedance to determine the frequency at which it vanishes. Thus the other two resonant frequencies $\omega_2$ and $\omega_3$ are found to be
\begin{align}
\omega_2 &=\omega_0(1-C_m/\sqrt{2}C_0),\\
\omega_3 &=\omega_0(1+C_m/\sqrt{2}C_0).
\end{align}
Comparing the above two equations with the ones calculated using CMT, it is can be found that
\begin{equation}
\kappa=\frac{C_m}{2\sqrt{C_0C_r}}.
\end{equation}
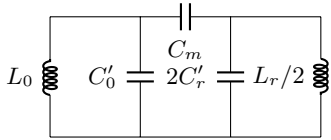
\begin{figure}[!h]
\begin{circuitikz}[scale=0.2]
\ctikzset {bipoles/length=0.6cm}
\draw
 (0,0) to [L, l^=$L_0$] (0,8)
 to (6,8)
 to [C, l_=$C_0'$] (6,0)
 (6,8) to [C, l_=$C_m$] (12,8) 
 to [C, l_=$2C_r'$] (12,0) (12,8) to (18,8)
 (18,8) to [L, l_=$L_r/2$] (18,0)
 (0,0) to (18,0);
\end{circuitikz}
\caption{The equivalent circuit of the unexcited system, when the symmetry plane T-T' is a Perfect Electric wall.}
\label{fig:CircuitModelAppendix3}
\end{figure}

\subsection*{State Space Modelling}
To derive a state space formulation for the three coupled centres, expressions for the time rate of the independent state variables (inductors currents and capacitors voltages) need to be obtained. For the circuit model in Fig. \ref{fig:CircuitModelAppendix1}, there are seven independent energy storage elements. Using basic circuit theory, the system of first ODEs can be written as
\begin{eqnarray*}
\left(
\begin{array}{c}
{dv_1}/{dt}\\
{dv_2}/{dt}
\end{array}
\right)&=
\left(
\begin{array}{cc}
C_0 & -C_m\\
-C_m & 2C_r
\end{array}
\right)^{-1}
\left(
\begin{array}{c}
i_S\\
-i_R
\end{array}
\right),\\
\left(
\begin{array}{c}
{dv_3}/{dt}\\
{dv_4}/{dt}
\end{array}
\right)&=
\left(
\begin{array}{cc}
2C_r & -C_m\\
-C_m & C_0
\end{array}
\right)^{-1}
\left(
\begin{array}{c}
i_R\\
-i_L
\end{array}
\right)
\end{eqnarray*}
for voltages across capacitors and
\begin{align*}
\frac{di_S}{dt}&=\frac{V_s}{L_0}-\frac{R_0+Rs}{L_0}i_S-\frac{v_1}{L_0},\\
\frac{di_R}{dt}&=\frac{1}{L_r}v_2-\frac{1}{L_r}v_3-\frac{R_r}{L_r}i_R,\\
\frac{di_L}{dt}&=\frac{1}{L_0}v_4-\frac{R_0+R_L}{L_0}i_L
\end{align*}
for currents through inductors.
\bibliography{EITref}
\end{document}